\documentclass{amsart}

\usepackage{gradient-benchmark}
\usepackage{amsmath}
\usepackage{amsfonts}
\usepackage{amssymb}
\usepackage{amsthm}
\usepackage{graphicx}
\usepackage{xr-hyper}
\usepackage[hyphens]{url}
\usepackage[hidelinks]{hyperref}
\usepackage{natbib}

\newcommand{\notarxiv}[1]{}
\newcommand{\arxiv}[1]{#1}

\notarxiv{
\jshort{mst}

\volname{}

\jvolume{0}

\jvol{}

\jissue{0}

\pubyear{2013}

\mstype{Article}

\artid{012}

\access{Advance Access publication March 3, 2013}
}

\begin{document}

\title[Phylogenetic gradient benchmark]{Automatic differentiation is no panacea for phylogenetic gradient computation}

\author[Fourment et al.]{
Mathieu Fourment,$^{1}$
Christiaan J Swanepoel,$^{2,3}$
Jared G Galloway,$^{4}$
Xiang Ji,$^{5}$
Karthik Gangavarapu,$^{6}$
Marc A Suchard,$^{\ast,6,7,8}$
Frederick A Matsen IV$^{\ast,4,9,10,11}$
}




\arxiv{
\maketitle

 \noindent{\small\it
 \address{
$^{1}$Australian Institute for Microbiology and Infection, University of Technology Sydney, Ultimo NSW, Australia;\\
$^{2}$Centre for Computational Evolution, The University of Auckland, Auckland, New Zealand;\\
$^{3}$School of Computer Science, The University of Auckland, Auckland, New Zealand, 1010;\\
$^{4}$Public Health Sciences Division, Fred Hutchinson Cancer Research Center, Seattle, Washington, USA;\\
$^{5}$Department of Mathematics, Tulane University, New Orleans, USA;\\
$^{6}$Department of Human Genetics, University of California, Los Angeles, USA;\\
$^{7}$Department of Computational Medicine, University of California, Los Angeles, USA;\\
$^{8}$Department of Biostatistics, University of California, Los Angeles, USA;\\
$^{9}$Department of Statistics, University of Washington, Seattle, USA;\\
$^{10}$Department of Genome Sciences, University of Washington, Seattle, USA;\\
$^{11}$Howard Hughes Medical Institute, Fred Hutchinson Cancer Research Center, Seattle, Washington, USA;\\
}
}

\medskip
 \noindent{\bf Corresponding authors:}
 Marc A. Suchard, Departments of Biomathematics, Biostatistics and Human Genetics, 6558 Gonda Building, 695 Charles E. Young Drive, South Los Angeles, CA 90095-1766, USA, E-mail: msuchard@ucla.edu;
 Frederick A. Matsen~IV, Fred Hutchinson Cancer Research Center, 1100 Fairview Ave.\ N, Mail stop S2-140, Seattle, WA, 98109, USA, E-mail: matsen@fredhutch.org\\
}

\notarxiv{
 \address{$^{1}$Australian Institute for Microbiology and Infection, University of Technology Sydney, Ultimo NSW, Australia;\\
$^{2}$Centre for Computational Evolution, The University of Auckland, Auckland, New Zealand;\\
$^{3}$School of Computer Science, The University of Auckland, Auckland, New Zealand, 1010;\\
$^{4}$Public Health Sciences Division, Fred Hutchinson Cancer Research Center, Seattle, Washington, USA;\\
$^{5}$Department of Mathematics, Tulane University, New Orleans, USA;\\
$^{6}$Department of Human Genetics, University of California, Los Angeles, USA;\\
$^{7}$Department of Computational Medicine, University of California, Los Angeles, USA;\\
$^{8}$Department of Biostatistics, University of California, Los Angeles, USA;\\
$^{9}$Department of Statistics, University of Washington, Seattle, USA;\\
$^{10}$Department of Genome Sciences, University of Washington, Seattle, USA;\\
$^{11}$Howard Hughes Medical Institute, Fred Hutchinson Cancer Research Center, Seattle, Washington, USA;\\
}

\history{Received 13 July 2011; reviews returned 26 November 2012; accepted 30 November 2012}

\coresp{E-mail: msuchard@ucla.edu; matsen@fredhutch.org
}
\datade{The POPRES data were obtained from dbGaP (accession no. phs000145.v1.p1).}

\editor{Robb Brumfield}
}

\begin{abstract}
Gradients of probabilistic model likelihoods with respect to their parameters are essential for modern computational statistics and machine learning.
These calculations are readily available for arbitrary models via ``automatic differentiation'' implemented in general-purpose machine-learning libraries such as TensorFlow and PyTorch.
Although these libraries are highly optimized, it is not clear if their general-purpose nature will limit their algorithmic complexity or implementation speed for the phylogenetic case compared to phylogenetics-specific code.
In this paper, we compare six gradient implementations of the phylogenetic likelihood functions, in isolation and also as part of a variational inference procedure.
We find that although automatic differentiation can scale approximately linearly in tree size, it is much slower than the carefully-implemented gradient calculation for tree likelihood and ratio transformation operations.
We conclude that a mixed approach combining phylogenetic libraries with machine learning libraries will provide the optimal combination of speed and model flexibility moving forward.
\end{abstract}

\notarxiv{
\maketitle
}

\arxiv{
}

\section*{Significance statement}
Bayesian phylogenetic analysis plays an essential role in understanding how organisms evolve, and is widely used as a tool for genomic surveillance and epidemiology studies.
The classical Markov chain Monte Carlo algorithm is the engine of most Bayesian phylogenetic software, however, it becomes impractical when dealing with large datasets.
To address this issue, more efficient methods leverage gradient information, albeit at the cost of increased computational demands.
Here we present a benchmark comparing the efficiency of automatic differentiation implemented in general-purpose libraries against analytical gradients implemented in specialized phylogenetic tools.
Our findings indicate that implementing analytical gradients for the computationally intensive components of the phylogenetic model significantly enhances the efficiency of the inference algorithm.

\section*{Introduction}
Gradients (i.e.\ multidimensional derivatives) of probabilistic model likelihoods with respect to their unknown parameters are essential for modern computational statistics and machine learning.
For example, gradient-based Hamiltonian Monte Carlo (HMC)~\citep{Neal2011-yo}, implemented in the Stan statistical framework~\citep{Carpenter2017-ui}, is a cornerstone of the modern Bayesian statistical toolbox.
Variational Bayesian (VB) inference algorithms~\citep{Blei2017-fs}, which use gradients to improve fit of a variational distribution to the posterior, are another key modern technique.
In the more general setting of machine learning, gradients are used to train predictive models such as deep neural networks.

Although gradients have been considered for a long time in phylogenetics~\citep{Schadt1998-zh,Kenney2012-zj}, they are now becoming of central importance to enable faster approaches to Bayesian phylogenetic analysis.
Bayesian methods have gained popularity among phylogenetic practitioners due to their ability to integrate multiple data sources, including ecological factors \citep{lemey2020accommodating} and clinical outcomes \citep{bedford2014integrating} into a single analysis.
A drawback of these methods is scalability, as it is well known that Bayesian phylogenetic packages, such as BEAST \citep{suchard2018bayesian}, struggle with datasets containing thousands of sequences with moderately complex models.
Bayesian phylogenetic analysis typically uses classical Markov chain Monte Carlo (MCMC) and therefore does not need to calculate computationally-intensive gradients.

In order to go beyond classical MCMC, recent research has developed Hamiltonian Monte Carlo~\citep{Fisher2021-mz} and Variational Bayes phylogenetic analysis~\citep{Fourment2019-fh,Zhang2019-lw,Dang2019-wq,Liu2021-lw,Moretti2021-ah,Ki2022-nt,Koptagel2022-bk,Zhang2022-tr}.
These methods require fast and efficient gradient calculation algorithms to give viable alternatives to MCMC\@.
Correspondingly, recent work has developed fast algorithms and implementations of phylogenetic likelihood gradient calculation~\citep{Ji2020-pp} in the \BEAGLE\ \citep{Ayres2019-rw} library.

Outside of phylogenetics, gradient-based analysis has also exploded in popularity, in part driven by easy to use software libraries that provide gradients via automatic differentiation (AD).
AD libraries ``record'' function compositions, have gradients on hand for component functions, and combine these simple gradients together via the chain rule (see~\citep{Margossian2019-xa} for a review).
This work has, remarkably, been extended to many computable operations that are not obviously differentiable such as dynamic control flow and unbounded iteration \citep{yu2018dynamic}.
These libraries, exemplified by TensorFlow~\citep{Abadi2016-xh} and PyTorch~\citep{Paszke2019-di}, are often developed by large dedicated teams of professional programmers.

The combination of these various advances raises a number of questions.
Can we rely on AD exclusively in phylogenetics, and avoid calculating gradients using hand-crafted algorithms?
How do AD algorithms scale when presented with interdependent calculations on a tree?
Does performance depend on the package used?

In this paper, we address these questions by performing the first benchmark analysis of AD versus carefully-implemented gradient algorithms in compiled languages.
We find that AD algorithms vary widely in performance depending on the backend library, the dataset size and the model/function under consideration.
All of these AD implementations are categorically slower than libraries designed specifically for phylogenetics; we do, however, find that they appear to scale roughly linearly in tree size.
Moving forward, these results suggest an architecture in which core phylogenetic likelihood and branch-length transformation calculations are performed in specialized libraries, whereas rich models are formulated, and differentiated, in a machine learning library such as PyTorch or TensorFlow.

\section*{Results}

\subsection*{Overview of benchmarking setup}
To coherently describe our results, we first provide a succinct overview of the phylogenetic and machine learning packages that we will benchmark as well as the computational tasks involved.

We benchmark two packages where the core algorithm implementation is specialized to phylogenetics: \BEAGLE~\citep{Ji2020-pp}, wrapped by our Python-interface C++ library \bito, as well as \physher~\citep{Fourment2014-sa}.
The \bito{} library also efficiently implements gradients of the ratio transformation, following~\citep{Ji2021-hc}, for unconstrained node-height optimization.
We compare these to the most popular AD libraries available, namely TensorFlow~\citep{Abadi2016-xh}, PyTorch~\citep{Paszke2019-di}, JAX~\citep{Bradbury2018-jax}, and Stan~\citep{Carpenter2017-ui}.
These are leveraged in phylogenetics via \treeflow, \torchtree, \phylojax, and \phylostan\ \citep{Fourment2019-fh} respectively.
When using AD, these programs make use of reverse-mode automatic differentiation.
Every program uses double precision unless specified otherwise.

We divide the benchmarking into two flavors: a ``micro-'' and ``macro-'' benchmark.
The macrobenchmark is meant to mimic running an actual inference algorithm, though stripped down to reduce the burden of implementing a complex model in each framework.
Specifically, we infer parameters of a constant-size coalescent process, strict clock, as well as node heights under a typical continuous-time Markov chain (CTMC) model for character substitution along an unknown phylogeny.
Every implementation uses the automatic differentiation variational inference (ADVI) framework~\citep{Kucukelbir2017-cr} to maximize the evidence lower bound (ELBO) over 5000 iterations.
\textit{A priori} we assume the CTMC substitution rate is exponentially distributed with mean 0.001 and we use the Jeffrey's prior for the unknown population size parameter.

The microbenchmark, on the other hand, is meant to identify which parts of a phylogenetic model are the most computationally expensive in the context of gradient-based inference.
This involves evaluating likelihoods and functions used in phylogenetic analysis and calculating their gradient (1) the phylogenetic likelihood, (2) the coalescent likelihood, (3) node height transform, and (4) the determinant of the Jacobian of the node height transform.
Specifically, these tasks are:

\begin{enumerate}
\item \textbf{Phylogenetic likelihood:} the likelihood of observing an alignment under the Jukes-Cantor substitution model~\citep{Jukes1969-xh} is efficiently calculated using the pruning algorithm~\citep{Felsenstein1981-zs} requiring $\mathcal{O}(N)$ operations where $N$ is the number of taxa.
In this benchmark, the derivatives are taken with respect to the branch lengths.
Although a naive implementation of the gradient calculation requires $\mathcal{O}(N^2)$ calculations, efficient implementations~\citep{Fourment2014-sa, Ji2020-pp} necessitate only $\mathcal{O}(N)$ operations.
We also benchmark the tree likelihood using the GTR substitution model.
The gradient with respect to the GTR parameters is calculated analytically in \physher\  while \bito\ utilizes finite differences.
Analytical gradients of the tree likelihood require $\mathcal{O}(N)$ operations for each of the 8 free parameters while numerical gradients require two evaluations of the tree likelihood per parameter.
\item \textbf{Coalescent likelihood:} the likelihood of observing a phylogeny is calculated using the constant size population coalescent model~\citep{Kingman1982-ag}.
The gradient with respect to the node heights and the population size parameter requires $\mathcal{O}(N)$ time.
\item \textbf{Node height transform:} Node ages of time trees need to be reparameterized in order to perform unconstrained optimization~\citep{Fourment2014-sa,Ji2021-hc}.
Evaluating this function requires a single preorder traversal and requires $\mathcal{O}(N)$ operations.
\item \textbf{Determinant of the Jacobian of the node height transform:} The transformation of the node ages requires an adjustment to the joint density through the inclusion of the determinant of the Jacobian of the transform~\citep{Fourment2019-fh}.
The Jacobian is triangular and the determinant is therefore straightforward to compute.
Although calculating its gradient analytically is not trivial, requiring $\mathcal{O}(N^2)$ calculations, recent work~\citep{Ji2021-hc} proposed an $\mathcal{O}(N)$ algorithm.
The derivatives are taken with respect to the node heights.
\end{enumerate}

\subsection*{AD implementations vary widely in performance, and custom gradients are far faster}

\begin{figure*}[h]
\centering
\includegraphics[width=0.95\textwidth]{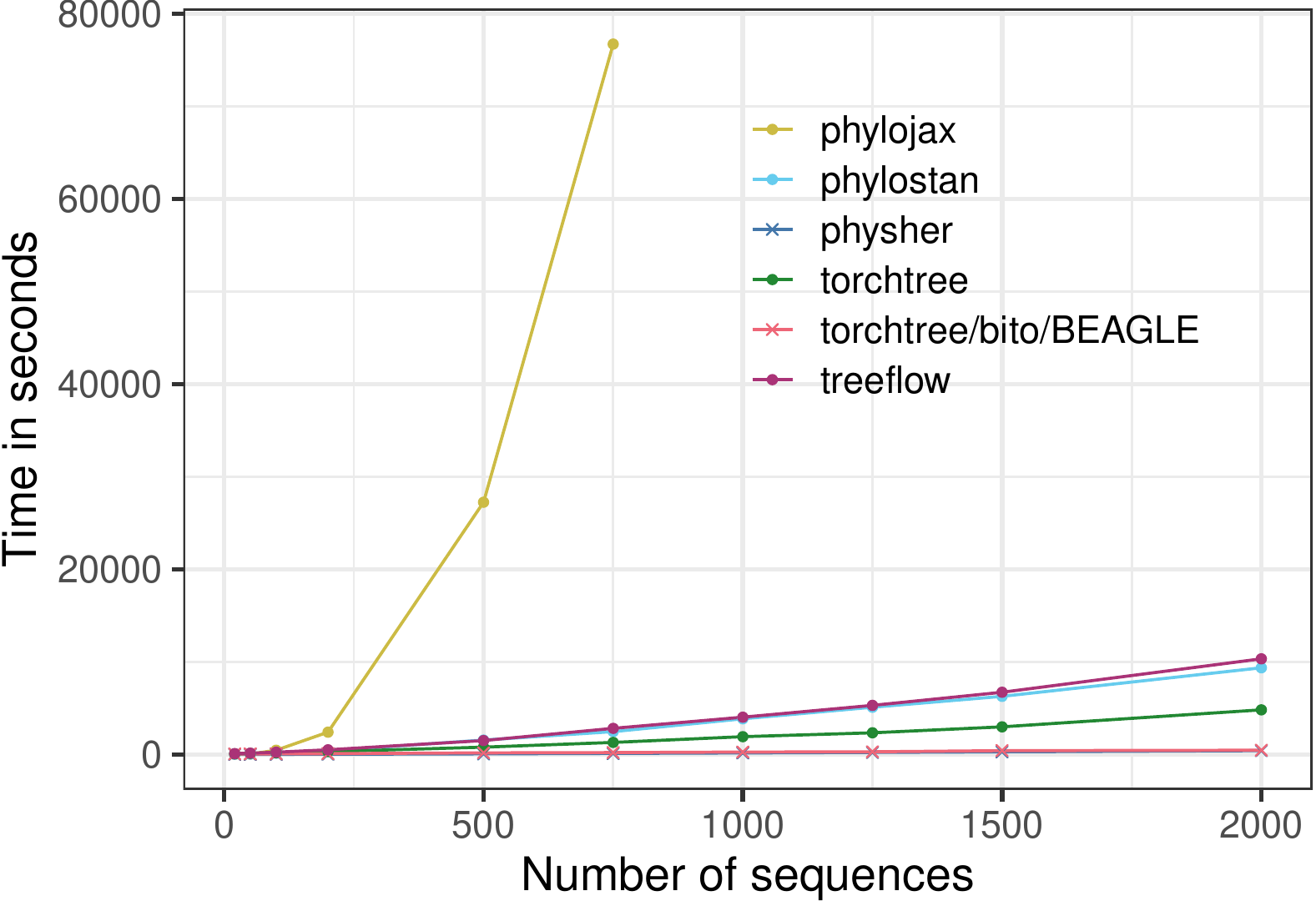}
\caption{\
Speed of implementations for 5000 iterations of variational time-tree inference with a strict clock.
See Figure~\ref{fig:macro_no_jaxl} for results without \phylojax.
}%
\label{fig:macro}
\end{figure*}

We find that on the macrobenchmark, AD implementations vary widely in their speed (Figure~\ref{fig:macro}).
This is remarkable given that these are highly optimized libraries doing the same flavor of operations.
Specifically, both just-in-time (JIT) compiled JAX and compiled TensorFlow use XLA as a backend, although they have strikingly different performance.%
\footnote{We note that this is now a known issue with JAX \url{https://github.com/google/jax/issues/10197}.}
Specifically, JAX was the only package that clearly scales quadratically in the number of tips.
Moreover, PyTorch was several times faster than TensorFlow for our tasks of interest, which was surprising to us because of PyTorch uses a dynamic computation graph.
Results for \phylojax\ with datasets larger than 750 sequences are not reported as they exceeded the maximum allocated computation time.

None of these AD libraries approach the speed of hand-coded phylogenetic gradients.
The BEAGLE gradients wrapped in \bito\ and gradients computed in \physher\ show comparable performance, which are at least 8 times the speed of the fastest AD implementation (Figure~\ref{fig:macro_time_rel}).

As expected, memory usage of the pure C program \physher\ is the smallest, while \torchtree\ is less memory heavy than \treeflow\ and \phylostan{}'s memory usage increases significantly more rapidly (Figure~\ref{fig:macro_mem}).
It is worth noting that \bito\ noticeably decreases the memory usage of \torchtree.

Overall using a specialized library for the tree likelihood within a Python program greatly improves the performance of a program making use of gradient-based optimization (e.g. ADVI, HMC) while incurring a small performance and memory cost compared to a fully C-based tool.

\subsection*{Relative performance of AD depends on the task}

\begin{figure*}[h]
\centering
\includegraphics[width=0.95\textwidth]{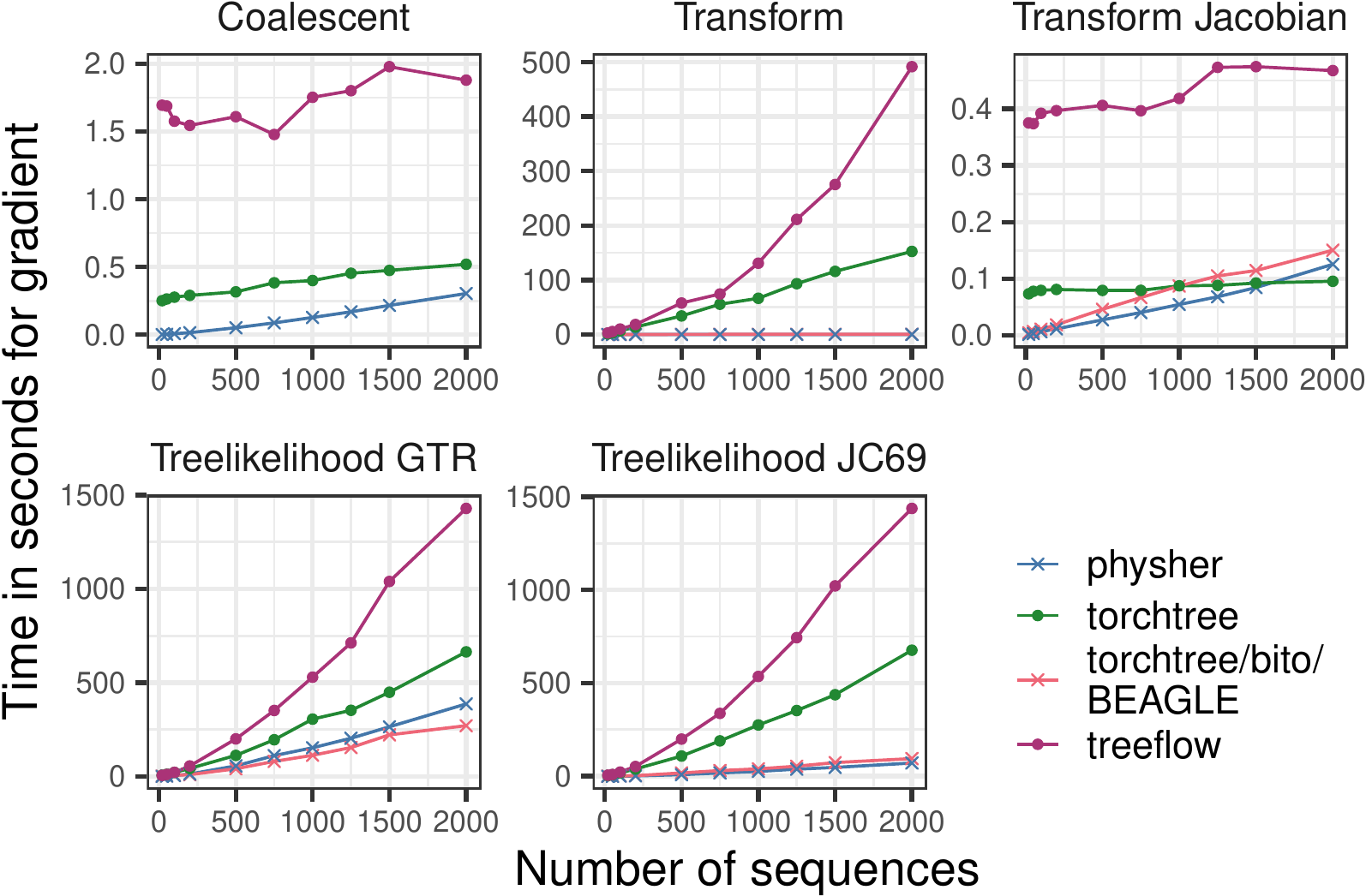}
\caption{\
Speed of implementations for the gradient of various tasks needed for inference.
See text for description of the tasks.
JAX is excluded from this plot due to slow performance stretching the $y$-axis; see Figure~\ref{fig:micro_grad_all} for JAX\@.
See Figure~\ref{fig:micro_eval_all} for function evaluations.
}%
\label{fig:micro}
\end{figure*}

To break down our inferential task into its components, we then performed a ``microbenchmark'' divided into the ingredients needed for doing gradient-based inference (Figure~\ref{fig:micro} and~\ref{fig:micro_eval}).
See Methods for a precise description of the individual tasks.
Across tasks, we see the following shared features.
The specialized phylogenetic packages (\bito/\BEAGLE\ and \physher) perform similarly to one another and are consistently faster than the AD packages, except for the Jacobian task.
As expected, the tree likelihood is the computational and memory bottleneck (Figure~\ref{fig:micro} and~\ref{fig:macro_mem})  in phylogenetic models and efficient gradient calculation are warranted.
TensorFlow-based \treeflow\ was the slowest implementation across the board after excluding JAX\@.

The AD programs also performed significantly worse in the node height transform and tree likelihood tasks.
Function calls in python are notoriously more expensive than in C and C++, potentially explaining the decrease in performance for algorithms involving a tree traversal.
In addition, the tree likelihood implementations in \BEAGLE\ and \physher\ are highly optimized with SSE vectorization \citep{Ayres2019-rw} and manual loop unrolling.

The calculations of the coalescent function and its gradient were slightly faster in  \physher\ than in \torchtree, although the difference was slight.
The ratio transform has nontrivial computational expense --- comparable to the phylogenetic likelihood gradient --- in AD packages; however, specialized algorithms for calculating these gradients scale much better.
Interestingly, for large datasets, \torchtree\ outperforms the specialized phylogenetic packages for the Jacobian ratio transform gradient calculation.
Since this is the fastest task, the overall execution time is not, however, significantly impacted.

The phylogenetic gradient is approximately linear for packages other than JAX (Figure~\ref{fig:micro_loglog}), although the specialized phylogenetic packages are about 10 times faster.
For the GTR calculation we actually compare two flavors of evaluation: finite differences for \bito\ and analytic gradients for \physher.
As expected, \bito\ is increasingly faster than \physher\ as the datasets increase in size.

With the exception of the tree likelihood, JAX's JIT capabilities greatly improved the performance of the algorithms in the microbenchmark (Figure~\ref{fig:micro_jax_jit}).
Analytically calculating the gradient of the tree likelihood considerably improved the running time of \phylojax\, pointing at implementation issues in the gradient function in JAX for this type of algorithm (Figure~\ref{fig:micro_jax_jit}).
In contrast, enabling JIT in \torchtree\ showed no improvement and was not included in the results.
The calculation of the tree likelihood and its gradient were significantly slower using single precision for datasets larger than 500 sequences.
This is because \torchtree, like most phylogenetic programs, rescales partial likelihood vectors in order to avoid underflow; using single precision requires more rescaling operations.

\section*{Discussion}

We have found that, although AD packages provide unrivaled flexibility for model development and flexible likelihood formulation, they cannot compete with carefully-implemented gradients in compiled languages.
Furthermore, they do differ between each other significantly in computation time and memory usage for phylogenetic tasks.

Our results motivate the design of \bito: leverage specialized algorithms for phylogenetic gradients and ratio transforms, but wrap them in a way that invites model flexibility.
In this paper, we have focused on two functionalities of \bito: first as a wrapper for the high-performance \BEAGLE\ library, and second, as a fast means of computing the ratio transforms.
This is our first publication using this library, which will be the computational core of our future work on Bayesian phylogenetic inference via optimization.
We will defer a more comprehensive description of \bito\ to future work.

Our results also motivate us to focus our future model developments using the PyTorch library, which shows the best performance as well as ease of use.

Our study has the following limitations.
First, these libraries are developing quickly and they may gain substantially in efficiency in future versions.
Second, these results concern CPU computation only.
Future work, including development of phylogenetic gradients using graphics processing units (GPUs), will evaluate the promise of GPUs for gradient-based inference.
However, we note that initial results using GPUs for AD packages did not lead to a significant speedup.

\section*{Methods}

\subsection*{Data}

To evaluate the performance of each implementation we reused parts of the validation workflow introduced by~\citep{Sagulenko2018-bb}.
The data in this workflow consists of a collection of influenza A datasets ranging from 20 to 2000 sequences sampled from 2011 to 2013.
Our benchmark is built on top of this pipeline and makes use of a reproducible Nextflow~\citep{DiTommaso2017-gy} pipeline.

\subsection*{Software benchmarked}
\torchtree\ is a Python-based tool that leverages the Pytorch library to calculate gradients using reverse mode AD.

\texttt{torchtree-bito} is a \torchtree\ plugin that offers an interface to the \bito\ library (\url{https://github.com/phylovi/bito}).
Within \bito{}, analytical derivatives with respect to the branch lengths are calculated through the \BEAGLE\ library \citep{Ayres2019-rw,Ji2020-pp} while the gradient with respect to the GTR substitution model parameters are calculated numerically using finite differences.
\bito\ and \BEAGLE\ do not provide analytical derivatives of the coalescent function, hence no results are shown in Figures~\ref{fig:micro} and \ref{fig:micro_eval}-\ref{fig:micro_loglog}.

\physher\ is a C program that allows one to approximate distributions using ADVI~\citep{Fourment2020-qw}, while every derivative is calculated analytically.
The derivatives with respect to the branch lengths are efficiently calculated using a linear-time algorithm developed independently of \citet{Ji2020-pp}.
The gradient of the Jacobian transform is efficiently calculated using the method proposed by \citet{Ji2021-hc}.

\phylostan\ is a Python-based program \citep{Fourment2019-fh} that generates phylogenetic models that are compatible with the Stan package.

\phylojax\ is a Python-based tool that leverages the JAX library to calculate gradients using reverse mode AD\@.

\treeflow\ is a Python-based tool that leverages the TensorFlow library to calculate gradients using reverse mode AD\@.
\treeflow's implementation of the phylogenetic likelihood uses TensorFlow's TensorArray construct \citep{yu2018dynamic}, a data structure which represents a collection of arrays. Each array can only be written once in a computation, and read many times. Using this data structure to implement the dynamic programming steps of the pruning algorithm potentially allows for more scalable gradient computations.

\begin{table}[h!]
\centering
\begin{tabular}{ |l|l|c| }
 \hline
Program & Availability & Version \\
\hline
\bito & \url{https://github.com/phylovi/bito} & autodiff-benchmark \\
\phylojax & \url{https://github.com/4ment/phylojax} & v1.0.1 \\
\phylostan & \url{https://github.com/4ment/phylostan} & v1.0.5 \\
\physher & \url{https://github.com/4ment/physher} & v2.0.0 \\
\torchtree & \url{https://github.com/4ment/torchtree} & gradient-benchmark \\
\texttt{torchtree-bito} & \url{https://github.com/4ment/torchtree-bito} & gradient-benchmark \\
\treeflow & \url{https://github.com/christiaanjs/treeflow} & autodiff-benchmark \\
\hline
\end{tabular}
\caption{Code availability and version number of each phylogenetic program.
Version identifiers correspond to git tags.}
\label{table:1}
\end{table}

\subsection*{Computational infrastructure}
The automated workflow was run using the Fred Hutchinson \texttt{gizmo} scientific computing infrastructure.
A single node with 36 (2 sockets by 18 cores) Intel~\textregistered{} Xeon Gold 6254 CPU @ 3.10GHz cores was used for all individual processes in the pipeline.
A total of 48G RAM was allocated.
The node was running on Ubuntu 18.04.5 LTS (Bionic Beaver)
with Nextflow (version 22.04.3.5703) and Singularity (version 3.5.3) modules installed.

\subsection*{Data Availability}
The Nextflow pipeline is available from \url{https://github.com/4ment/gradient-benchmark}.
The versions of the programs used in this study are provided in Table~\ref{table:1}.

\section*{Acknowledgements}

We are grateful to Jonathan Terhorst for discussions concerning phylogenetic gradients in JAX\@.
This work was supported through US National Institutes of Health grants AI162611 and AI153044.
Scientific Computing Infrastructure at Fred Hutch was funded by ORIP grant S10OD028685.
Computational facilities were provided by the UTS eResearch High Performance Compute Facilities.
Dr.\ Matsen is an Investigator of the Howard Hughes Medical Institute.

\arxiv{
\bibliographystyle{plainnat}
}
\notarxiv{
\bibliographystyle{natbib}
}
\bibliography{main}

\newpage


\begin{center}
\textbf{Supplementary Material}
\end{center}

\renewcommand{\thefigure}{S\arabic{figure}}
\setcounter{page}{1}
\setcounter{section}{0}
\setcounter{table}{0}
\setcounter{figure}{0}
\setcounter{equation}{0}

\begin{figure*}[h]
\centering
\includegraphics[width=0.95\textwidth]{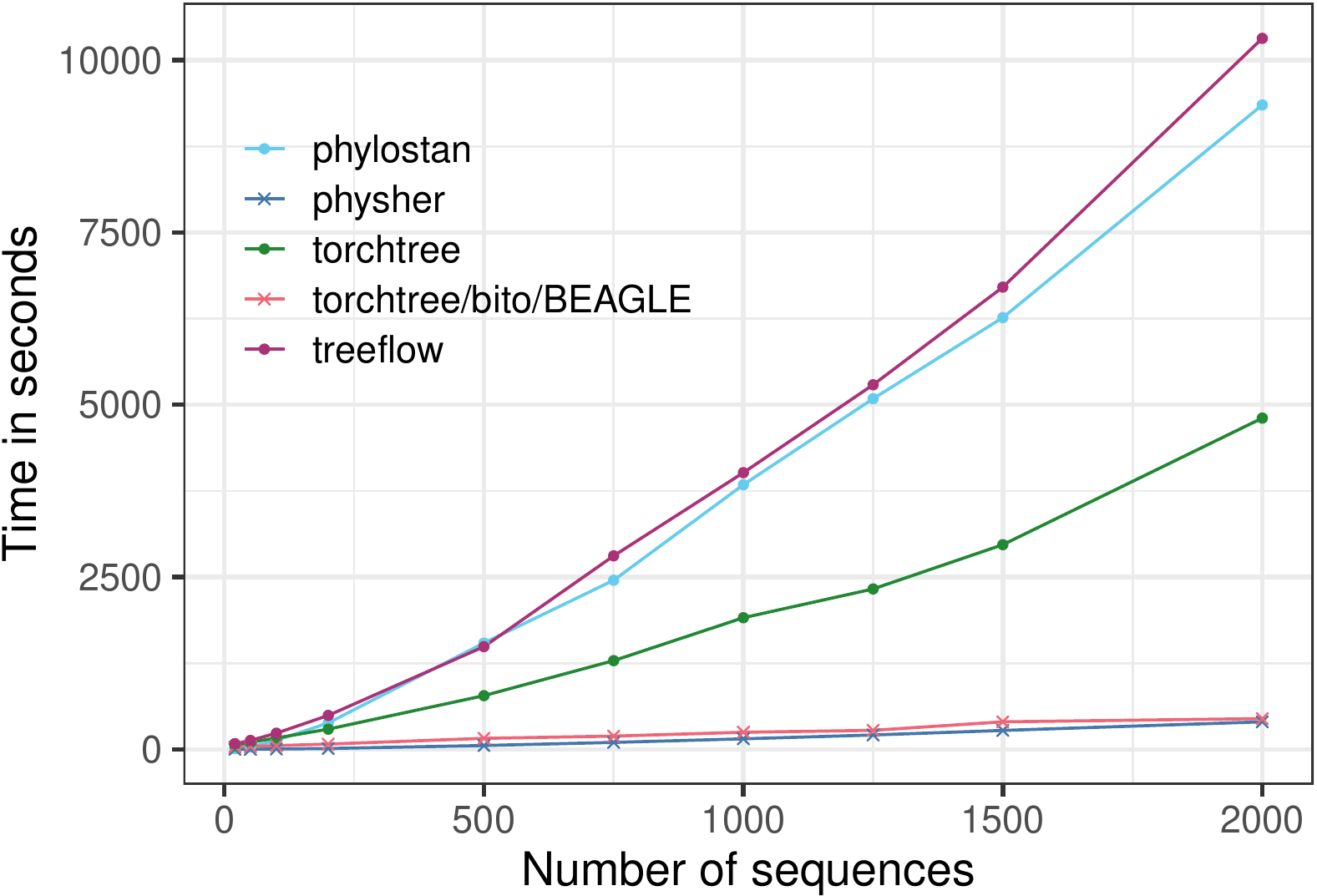}
\caption{\
Speed of implementations for 5000 iterations of variational time-tree inference with a strict clock.
}%
\label{fig:macro_no_jaxl}
\end{figure*}

\begin{figure*}[h]
\centering
\includegraphics[width=0.95\textwidth]{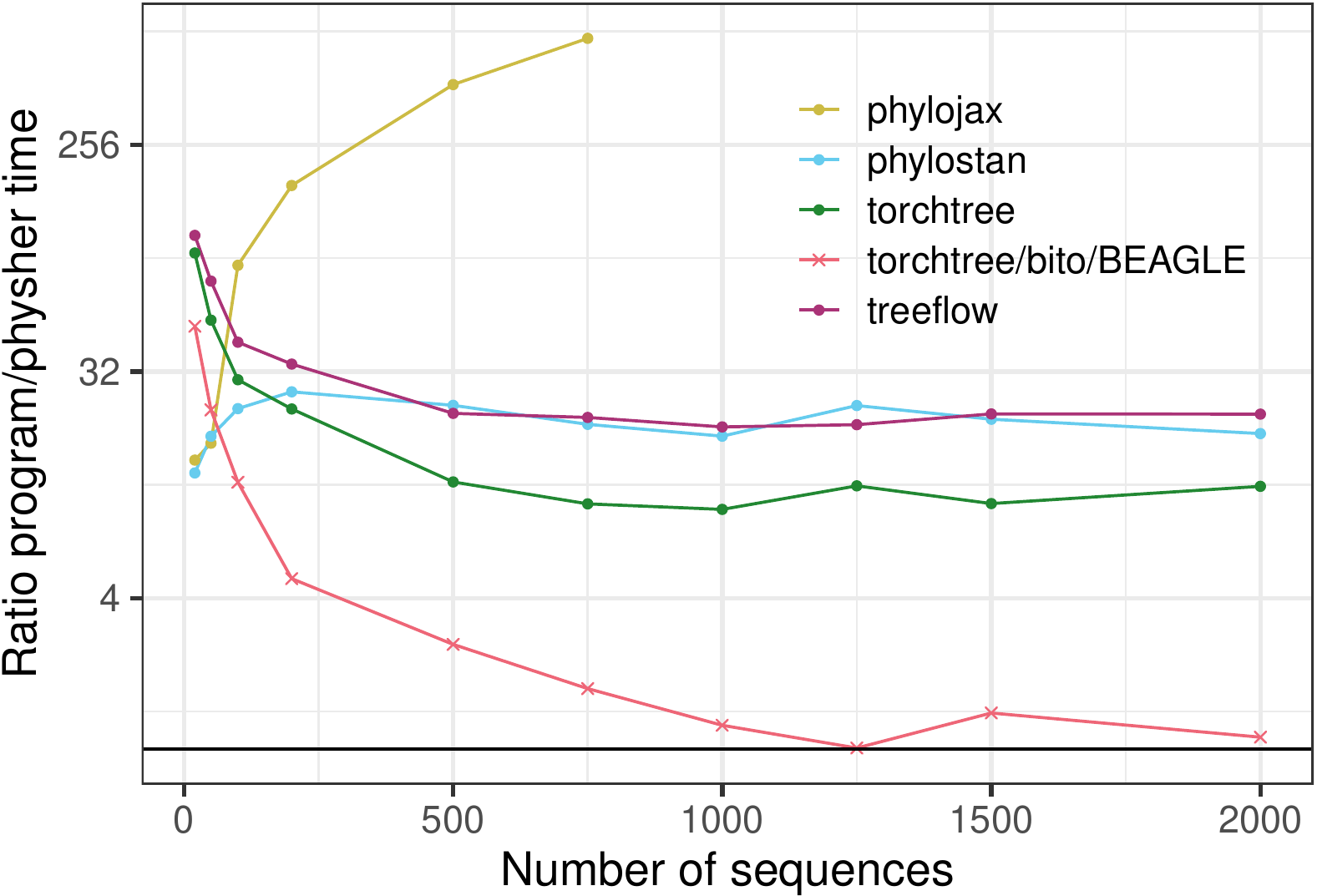}
\caption{\
Relative performance of each implementation against \physher.
The black horizontal line intersects the $y$-axis at 1.
}%
\label{fig:macro_time_rel}
\end{figure*}

\clearpage

\begin{figure*}[h]
\centering
\includegraphics[width=0.95\textwidth]{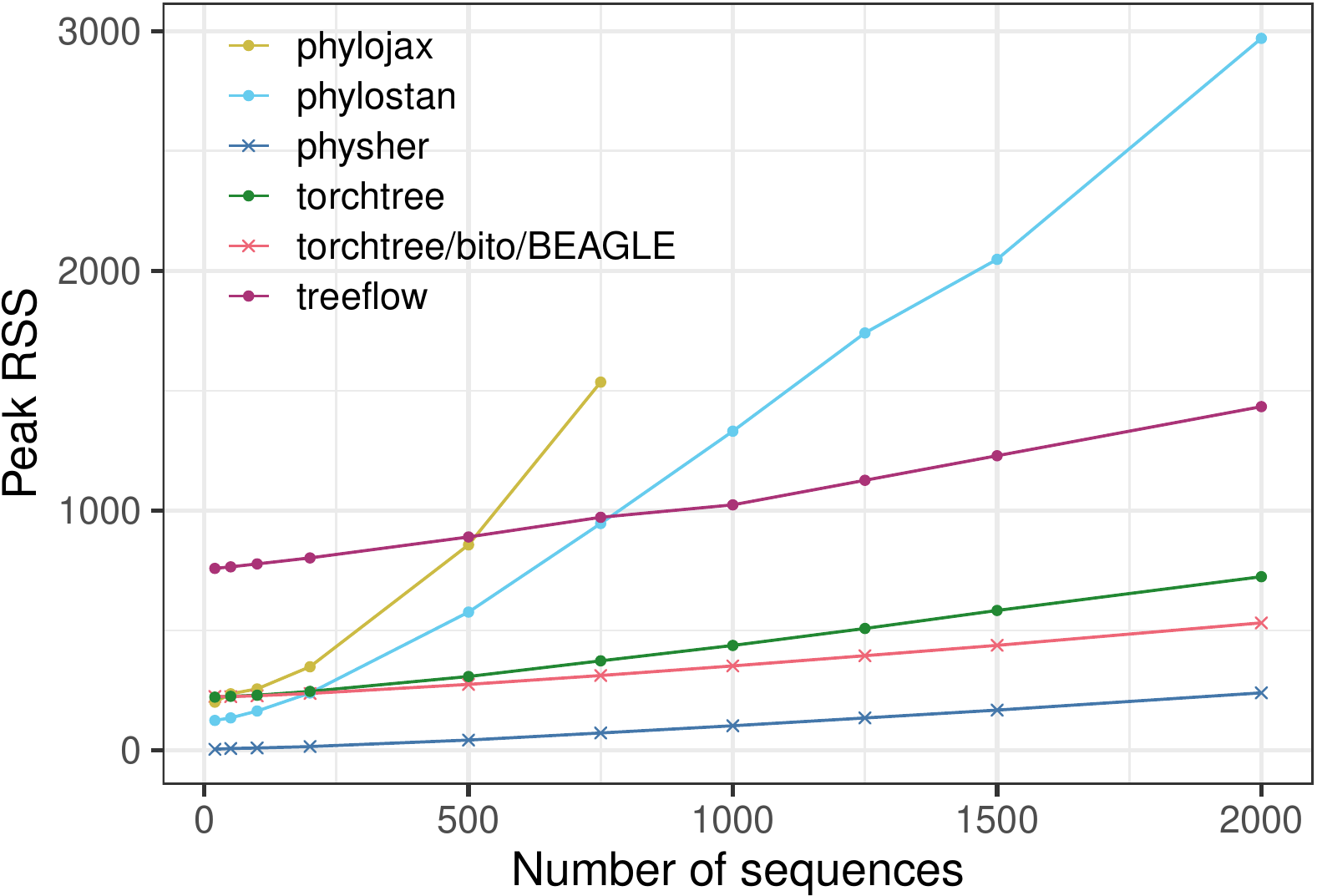}
\caption{\
Peak RSS (resident set size) memory usage of implementations for 5000 iterations of variational time-tree inference with a strict clock.
}%
\label{fig:macro_mem}
\end{figure*}

\begin{figure*}[h]
\centering
\includegraphics[width=0.95\textwidth]{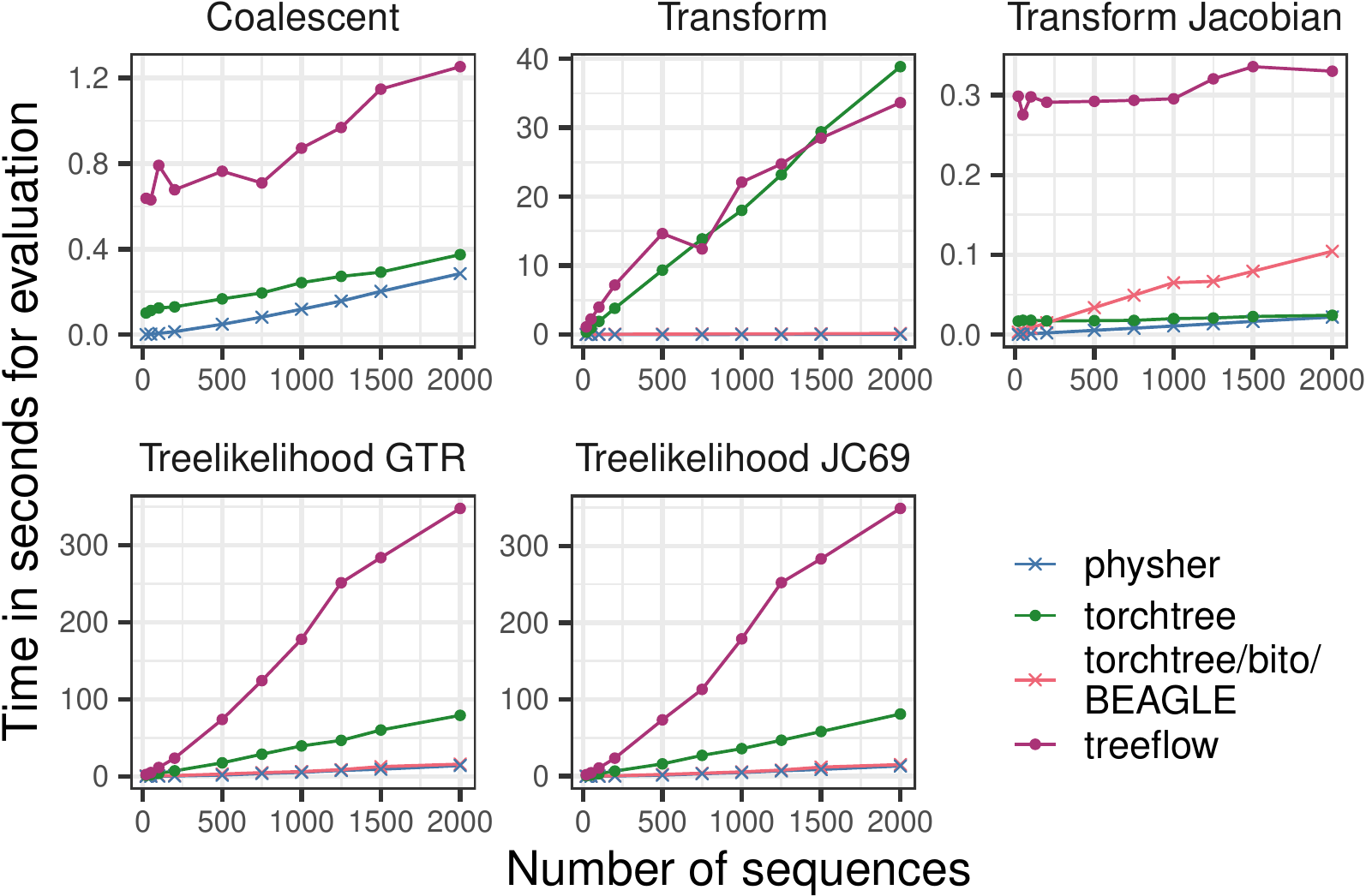}
\caption{\
Speed of implementations for the evaluation component for various inferential tasks.
See text for description of the tasks.
\phylojax\ results are excluded from this plot; see Figure~\ref{fig:micro_eval_all} for \phylojax.
See Figure~\ref{fig:micro_eval_all} for function evaluations.
}%
\label{fig:micro_eval}
\end{figure*}

\begin{figure*}[h]
\begin{center}
\includegraphics[width=0.95\linewidth]{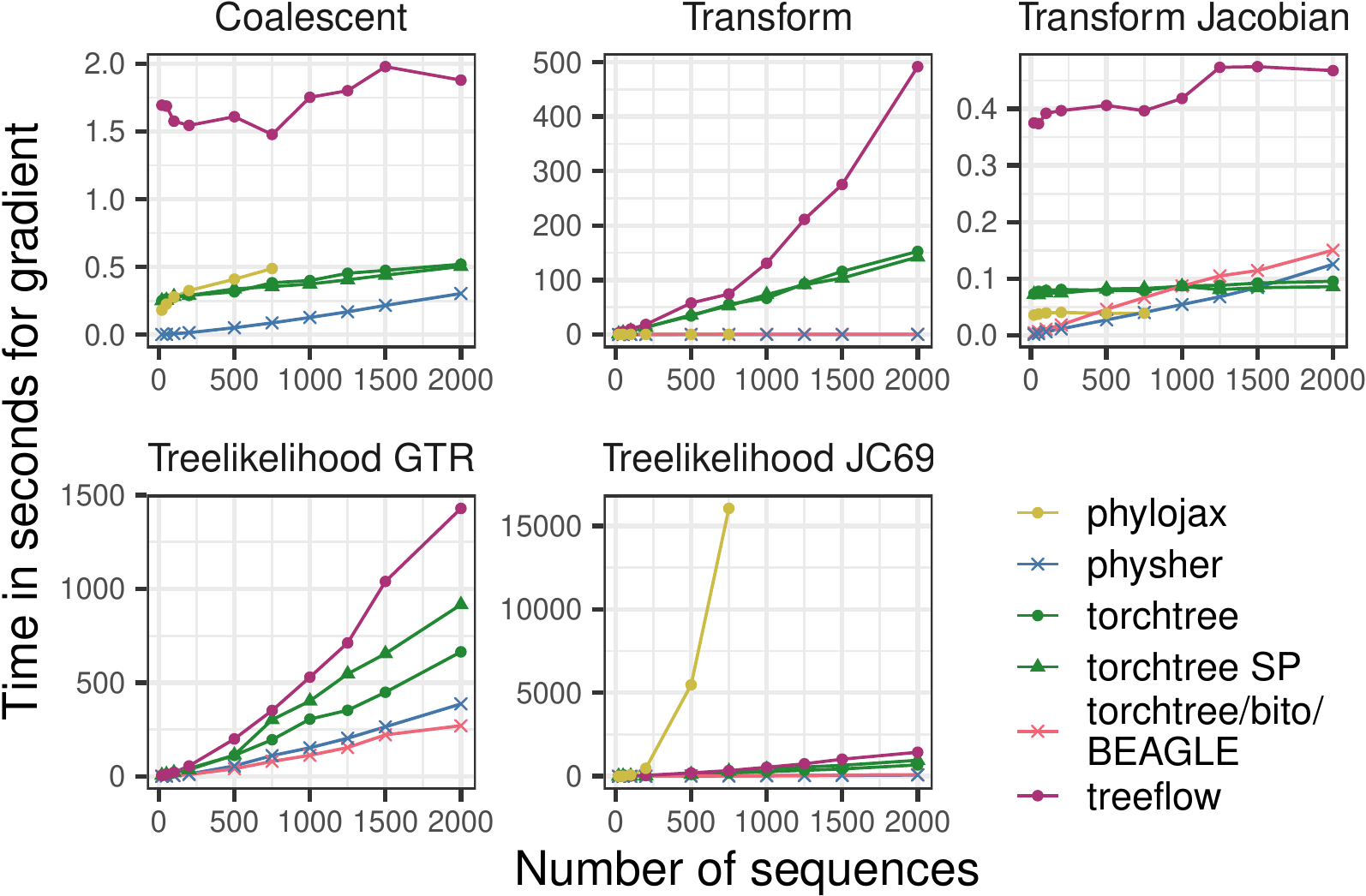}
\end{center}
\caption{Speed of implementations for the gradient component for various inferential tasks.
The \textsf{torchtree SP} label denotes \torchtree\ running with single precision.
Just-in-time compilation is enabled for \phylojax.}
\label{fig:micro_grad_all}
\end{figure*}

\begin{figure*}[h]
\begin{center}
\includegraphics[width=0.95\linewidth]{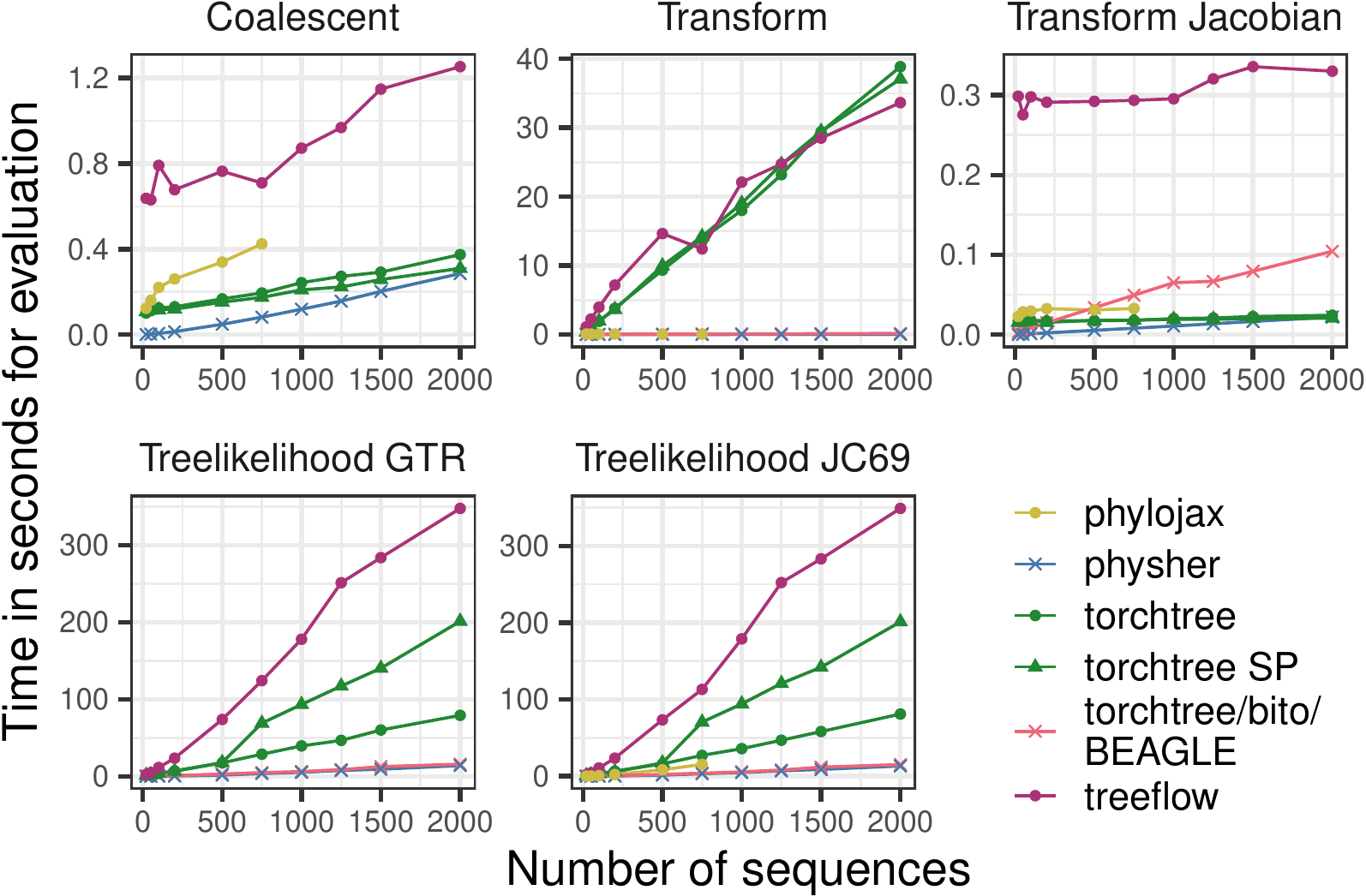}
\end{center}
\caption{Speed of implementations for the evaluation component for various inferential tasks.
The \textsf{torchtree SP} label denotes \torchtree\ running with single precision.
Just-in-time compilation is enabled for \phylojax.}
\label{fig:micro_eval_all}
\end{figure*}

\clearpage

\begin{figure*}[h]
\begin{center}
\includegraphics[width=0.95\linewidth]{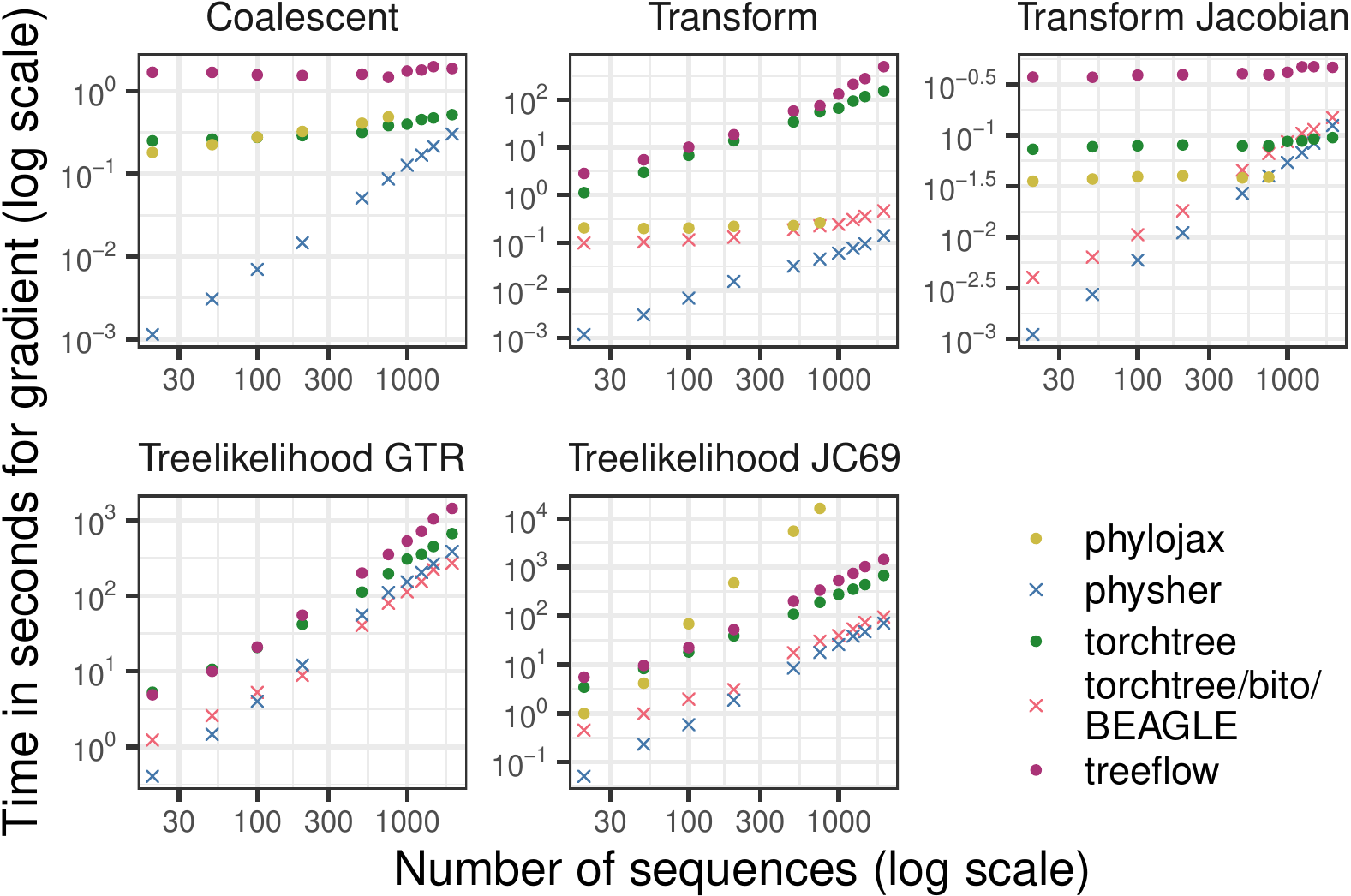}
\end{center}
\caption{Log-log plot of gradient calculation time against dataset size for various inferential tasks.
Just-in-time compilation is enabled for \phylojax.}
\label{fig:micro_loglog}
\end{figure*}

\begin{figure*}[h]
\begin{center}
\includegraphics[width=0.95\linewidth]{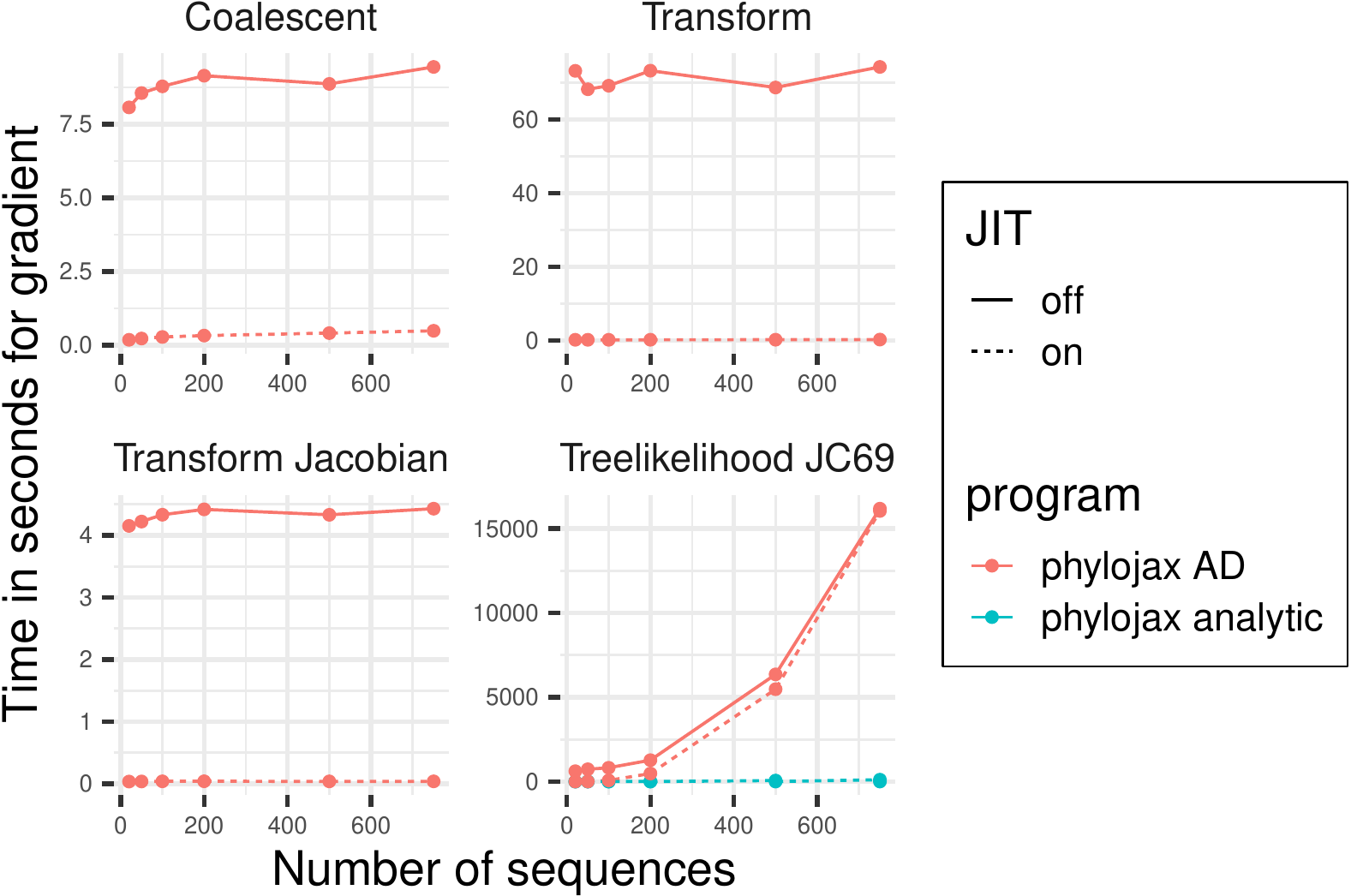}
\end{center}
\caption{Gradient calculation time against dataset size for various inferential tasks with \phylojax.
Gradient are calculated using automatic differentiation (AD) or finite differences.
Just-in-time (JIT) compilation is either turned on or off.}
\label{fig:micro_jax_jit}
\end{figure*}



\end{document}